\documentclass[10pt]{llncs}

\usepackage{pslatex}
\usepackage{colortbl}
\usepackage{graphicx}
\usepackage{amssymb}
\usepackage{url}
\usepackage{latexsym}
\usepackage{amsfonts}
\usepackage{amsmath}

\title{Toward a Formal Model of the Shifting Relationship between Concepts and Contexts during Associative Thought}
\author {Tomas Veloz$^1$ 
\and Liane Gabora$^1$
\and Mark Eyjolfson$^1$
\and Diederik Aerts$^2$} 
\authorrunning{Tomas Veloz et al.} 
 \titlerunning{Shifting Relationship between Concepts and Contexts}
\institute{University of British Columbia, 
Department of Psychology, Okanagan campus, $3333$ University Way
    Kelowna BC, V1V 1V7, CANADA,\\
    \email{tomas.veloz@ubc.ca, liane.gabora@ubc.ca, mark\_eyjolfson@hotmail.com},\and 
    Center Leo Apostel for Interdisciplinary Studies, Vrije Universiteit Brussel, Belgium,\\
    \email{diraerts@vub.ac.be}}

\begin{document}
\maketitle
\begin{abstract}
The quantum inspired State Context Property (SCOP) theory of concepts is unique amongst theories of concepts in offering a means of incorporating that for each concept in each different context there are an unlimited number of exemplars, or states, of varying degrees of typicality. Working with data from a study in which participants were asked to rate the typicality of exemplars of a concept for different contexts, and introducing an exemplar typicality threshold, we built a SCOP model of how states of a concept arise differently in associative versus analytic (or divergent and convergent) modes of thought. Introducing measures of state robustness and context relevance, we show that by varying the threshold, the relevance of different contexts changes, and seemingly atypical states can become typical. The formalism provides a pivotal step toward a formal explanation of creative thought processes.

\hspace{15 mm}


\hspace{15 mm}

\textbf{Reference:}
Veloz, T., Gabora, L., Eyjolfson, M., \& Aerts, D. (2011). A model of the shifting relationship between concepts and contexts in different modes of thought. Lecture Notes in Computer Science 7052: Proceedings of Fifth International Symposium on Quantum Interaction. June 27-29, Aberdeen, UK. Berlin: Springer.

\end{abstract}

\section{Introduction}

This paper unites two well-established psychological phenomena using a quantum-inspired mathematical theory of concepts, the State-COntext-Property (SCOP) theory of concepts. The first phenomenon is that the meaning of concepts shifts, sometimes radically, depending on the context in which they appear~\cite{Barsalou1982,Gabora2008,Hampton1997}. It is this phenomenon that SCOP was developed to account for~\cite{Gabora2002,Aerts2005a,Aerts2005b}. 
Here we use SCOP to model a different though related psychological phenomenon. This second psychological phenomenon was hinted at in the writings of a number of the pioneers of psychology, including Freud~\cite{Freud1949}, Piaget~\cite{Piaget1926}, and William James ~\cite{James1890}. They and others have suggested that all humans possess two distinct ways of thinking. The first, sometimes referred to as divergent or {\it associative thought}, is thought to be automatic, intuitive, diffuse, unconstrained, and conducive to unearthing remote or subtle associations between items that share features, or that are {\it correlated} but not necessarily {\it causally} related. This may yield a promising idea or solution though perhaps in a vague, unpolished form. There is evidence that associative thinking involves controlled access to, and integration of, affect-laden material, or what Freud referred to as ``primary process'' content~\cite{Freud1949,Russ1993}. Associative thought is contrasted with a more controlled, logical, rule-based, convergent, or {\it analytic} mode of thought that is conducive to analyzing relationships of cause and effect between items already believed to be related. Analytic thought is believed to be related to what Freud termed ``secondary process'' material.

A growing body of experimental and theoretical evidence for these two modes of thought, associative and  analytic, led to hypothesis that thought varies along a continuum between these two extremes depending on the situation we are in ~\cite{Ashby2002,Finke1992,Freud1949,Gabora2002,Gabora2003,Gabora2008,Guilford1950,James1890}. The capacity to shift between the two modes is sometimes referred to as {\it contextual focus}, since a change from one mode of thought to the  other is is brought about by the context, through the focusing or defocusing of attention~\cite{Gabora2003,Gabora2010}. Contextual focus is closely related to the dual-process theory of human cognition, the idea that human thought employs both implicit and explicit ways of learning and processing information~\cite{Chaiken1999,Evans2009}. It is not just the existence of two modes of thought but the cognitive consequences of shifting between them, that we use SCOP to model in this paper. 




\section{The SCOP Theory of Concepts}

The SCOP formalism is an operational approach in the foundations of quantum mechanics in which a physical system is determined by the mathematical structure of its set of states, set of properties and the possible (measurement) contexts which can be applied to this entity, and the relations between these sets. The SCOP formalism is part of a longstanding effort to develop an operational approach to quantum mechanics known as the Geneva-Brussels approach~\cite{Piron1976}. If a suitable set of quantum axioms is satisfied by the set of properties, one recovers via the Piron-Sol\`er
representation theorem the standard description of quantum mechanics in Hilbert space~\cite{Piron1976}. The SCOP formalism permits one to describe not only physical entities, but also potential entities 
~\cite{Aerts2002}, which means that SCOP aims at a very general description of how the interaction between context and the state of an entity plays a fundamental role in its evolution. In this work we make use of the SCOP formalism to model concepts, continuing the research reported in~\cite{Aerts2005a,Aerts2005b,Gabora2002,Gabora2009}.

Formally a conceptual SCOP entity consists of three sets $\Sigma$, ${\cal M}$, and ${\cal L}$: the set of states, the set of contexts and the set of properties, and two additional functions $\mu$ and $\nu$. The function $\mu$ is a probability function that describes how state $p$ under the influence of context $e$ changes to state $q$. Mathematically, this means that $\mu$ is a function from the set $\Sigma \times {\cal M} \times {\Sigma}$ to the interval $[0, 1]$, where $\mu(q, e, p)$ is the probability that state $p$ under the influence of context $e$ changes to state $q$. We write 
\begin{eqnarray}
\mu: \Sigma \times {\cal M} \times {\Sigma} &\rightarrow& [0, 1] \nonumber \\
(q, e, p) &\mapsto& \mu(q, e, p)
\end{eqnarray}
The function $\nu$ describes the weight, which is the renormalization of the applicability, of a certain property given a specific state. This means that $\nu$ is a function from the set $\Sigma \times {\cal L}$ to the interval $[0, 1]$, where $\nu(p, a)$ is the weight of property $a$ for the concept in state $p$. We write 
\begin{eqnarray}
\nu : \Sigma \times {\cal L} &\rightarrow& [0, 1] \nonumber \\
(p, a) &\mapsto& \nu(p, a)
\end{eqnarray}
Thus the SCOP is defined by the five elements $(\Sigma, {\cal M}, {\cal L}, \mu, \nu)$. States of a concept are denoted by means of the letters $p, q, r, \dots$ or $p_1, p_2, \ldots$, and contexts by means of the letters $e, f, g, \ldots$ or $e_1, e_2, \ldots$. When a concept is not influenced by any context, we say is in its {\it ground state}, and we denote the ground state by $\hat p$. The unit context, denoted $1$, is the absence of a specific context. Hence context $1$ leaves the ground state $\hat p$ unchanged. Exemplars of a concept are states of this concept in the SCOP formalism.

Note that in SCOP, concepts exist in what we refer to as a state of potentiality until they are evoked or actualized by some context. To avoid misunderstanding we mention that $\mu(p,e,q)$ is not a conditional probability of 
transitioning from state $p$ to $q$ {\it given that} the context is $e$. Contexts in SCOP are not just conditions, but active elements that {\it alter} the state of the concept, analogous to the {\it observer phenomenon} of quantum physics, where measurements affect the state of the observed entity. Indeed, a SCOP concept can be represented in a complex Hilbert space ${\cal H}$. Each state $p$ is modelled as a unitary vector (pure state) $|p\rangle\in {\cal H}$, or a trace-one density operator (density state) $\rho_p$. 
A context $e$ is generally represented by a linear operator of the Hilbert space ${\cal H}$, that provokes a probabilistic collapse by a set of orthogonal projections $\{P_i^e\}$. A property $a$ is always represented by an orthogonal projector $P_a$ in ${\cal H}$ respectively.
The contextual influence of a context on a concept is modelled by the application of the context operator on the concept's state. 
A more detailed explanation can be found in~\cite{Aerts2005a,Aerts2005b}.

\section{The Study}

Our application of SCOP made use of data obtained in a psychological study of the effect of context on the typicality of exemplars of a concept. We now describe the study. 

\subsection{Participants} 

Ninety-eight University of British Columbia undergraduates who were taking a first-year psychology course participated in the experiment. They received credit for their participation. 

\subsection{Method}

The study was carried out in a classroom setting. The participants were given questionnaires that listed eight exemplars (states) of the concept HAT. The exemplars are: state $p_1$: `Cowboy hat', state $p_2$: `Baseball cap', state $p_3$: `Helmet', state $p_4$: `Top hat', state $p_5$: `Coonskincap', state $p_6$: `Toque', state $p_7$: `Pylon', and state $p_8$: `Medicine hat'. They were also given five different contexts. The contexts are: the default or unit context $e_1$: {\it The hat}, context $e_2$: {\it Worn to be funny}, context $e_3$: {\it Worn for protection}, context $e_4$: {\it Worn in the south}, and context $e_5$: {\it Not worn by a person}.

The participants were asked to rate the typicality of each exemplar on a $7$-point Likert scale, where $0$ points represents ``not at all typical'' and $7$ points represents ``extremely typical''. Note that all the contexts except $e_1$ make reference to the verb ``wear'', which is relevant to the concept HAT. The context $e_1$ is included to measure the typicality of the concept in a context that simulates the {\it pure} meaning of a HAT, i.e. having no contextual influence, hence what in SCOP is meant by ``the unit context''. 

\subsection{Results}

A summary of the participants' ratings of the typicality of each exemplar of the concept HAT for each context is presented in Table 1. The contexts are shown across the top, and exemplars are given in the left-most column. For each pair $(a;b)$  in the table, $a$ represents the averaged sum of the Likert points across all participants. $b$ is the renormalized typicality of the state and context specified by the row and column respectively. The bottom row gives the normalization constant of each renormalized typicality function. Grey boxes have renormalized typicality below the threshold $\alpha=0.16$. 

\begin{table}
\begin{center}
\begin{tabular}{|l|c|c|c|c|c|}\hline
 {\it Exp. Data}	&$e_1$& $e_2$ & $e_3$ & $e_4$ & $e_5$ \\ \hline
$p_1$ Cowboy hat	& (5.44;0.18)	&\cellcolor[gray]{0.8}(3.57;0.14)&	\cellcolor[gray]{0.8}(3.06;0.13)&	(6.24;0.28)	&\cellcolor[gray]{0.8}(0.69;0.05)\\ \hline
$p_2$  Baseball cap	&(6.32;0.21)&	\cellcolor[gray]{0.8}(1.67;0.06)&	\cellcolor[gray]{0.8}(3.16;0.13)&	(4.83;0.21)	&\cellcolor[gray]{0.8}(0.64;0.04)\\ \hline
$p_3$ Helmet	&\cellcolor[gray]{0.8}(3.45;0.11)	&\cellcolor[gray]{0.8}(2.19;0.08)	&(6.85;0.28)	&\cellcolor[gray]{0.8}(2.85;0.13)&	\cellcolor[gray]{0.8}(0.86;0.06)\\ \hline
$p_4$  Top hat&	(5.12;0.17)&(4.52;0.17)&		\cellcolor[gray]{0.8}(2.00;0.08)&		\cellcolor[gray]{0.8}(2.81;0.12)&		\cellcolor[gray]{0.8}(0.92;0.06)\\ \hline
$p_5$ Coonskincap&\cellcolor[gray]{0.8}(3.55;0.11)	& (5.10;0.19)	&\cellcolor[gray]{0.8}(2.57;0.10)	&\cellcolor[gray]{0.8}(2.70;0.12)	&\cellcolor[gray]{0.8}(1.38;0.1)\\ \hline
$p_6$ Toque&	\cellcolor[gray]{0.8}(4.96;0.16)	&\cellcolor[gray]{0.8}(2.31;0.09)	&(4.11;0.17)	&\cellcolor[gray]{0.8}(1.52;0.07)	&\cellcolor[gray]{0.8}(0.77;0.05)\\ \hline
$p_7$ Pylon&\cellcolor[gray]{0.8}	(0.56;0.02)	&(5.46;0.21)	&\cellcolor[gray]{0.8}(1.36;0.05)	&\cellcolor[gray]{0.8}(0.68;0.03)&	(3.95;0.29)\\ \hline
$p_8$ Medicine hat&	\cellcolor[gray]{0.8}	(0.86;0.02)	&\cellcolor[gray]{0.8}	(1.14;0.04)&	\cellcolor[gray]{0.8}	(0.67;0.03)	&\cellcolor[gray]{0.8}	(0.56;0.02)&(4.25;0.31)\\ \hline
$T(e)$ & 30.30 &  25.98 & 23.80 &  22.22 & 13.51 \\ \hline
\end{tabular}
\vspace{0.5cm}
\caption{Summary of the participants' ratings of the typicality of the different exemplars of the concept HAT for different contexts.}
\label{data}
\end{center}
\vspace{-1cm}
\end{table}

\section{Analysis of Experimental Data and Application to the Model}
In this section we use SCOP to analyze the data collected in the experiment, and apply it to the development of a tentative formal model of how concepts are used differently in analytic and associative thought. 

\subsection{Assumptions and Goals}
We model the concept HAT with the SCOP $(\Sigma,{\cal M},{\cal L},\mu,\nu)$ where $\Sigma=\{p_1, \ldots, p_8\}$ and ${\cal M}=\{e_1, \ldots, e_5\}$ are the sets of exemplars and contexts considered in the experiment (see table~\ref{data}). We did not consider properties of the concept HAT, and hence ${\cal L}$ and $\nu$ are not specified. This is a small and idealized SCOP model, since only one experiment with a fairly limited number of states and contexts is considered, but it turned out to be sufficient to carry out the qualitative analysis we now present. Moreover, it will be clear that the approach can be extended in a straightforward way to the construction of more elaborate SCOP models that include the applicabilities of properties. Note also that the Hilbert space model of this SCOP can be constructed following the procedure explained in~\cite{Aerts2005b}. 

Recall how the participants estimated the typicality of a particular exemplar $p_i$, $i\in\{1, \ldots, 8\}$ under a specified context $e_j$, $j\in\{1, \dots, 5\}$ by rating this typicality from 0 to 7 on a Likert scale. Since these ratings play a key role in the analysis, we introduce the Likert function $L$:
\begin{eqnarray}
L: \Sigma \times {\cal M} &\rightarrow& [0, 7] \\
(p, e) &\mapsto& L(p,e)
\end{eqnarray}
where $L(p,e)$ is the Likert score averaged over all participants for state $p$ under context $e$. 

We also introduce the total Likert function $T$ which gives the total Likert score for a given context:
\begin{equation}
\begin{split}
T : {\cal M} &\rightarrow [0, 56] \\
e &\mapsto T(e)=\sum\limits_{p\in \Sigma} L(p,e),
\end{split}
\end{equation}
The Likert score $L(p,e)$ is not directly connected to the transition probability $\mu(p,e,\hat p)$ from the ground state of a concept to the state $p$ under context $e$. However, the renormalized value of $L(p,e)$ to the interval $[0,1]$ 
provides a reasonable estimate of the transition probability $\mu(p,e,\hat p)$. Hence we introduce the hypothesis that the renormalized Likert scores correspond to the transition probabilities from the ground state, or
\begin{equation}
\mu(p, e, \hat p)={L(p,e) \over T(e)}
\end{equation}
This is an idealization 
since the transition probabilities are independent although correlated to the renormalized Likert scores. 
In 
future work we plan experiments to directly measure the transition probabilities.

Let us pause briefly to explain why  these functions have been introduced. If we consider the unit context, it would be natural to link the typicality to just the Likert score. For example, for the unit context, exemplar $p_1$: `Cowboy hat' is more typical than $p_6$: `Toque' because $L(p_6,e_1) < L(p_1, e_1)$ (see table~\ref{data}). If one examines more than one context, however, such a conclusion cannot easily be drawn. For example, consider the exemplar $p_7$: `Pylon', under both the context $e_2$: {\it Worn to be funny} and context $e_5$: {\it Not worn by a person}, we have that $L(p_7, e_5) < L(p_7, e_2)$, but $p_7$ is more typical under context $e_5$ than under $e_2$. This is because $T(e_5)<T(e_2)$, i.e. the number of Likert points given in total for context $e_2$ is much higher than the number of Likert points given in total for the context $e_5$. This is primarily due to the fact that Likert points have been attributed by participant per context. 



Note that $\frac{T(e)}{8}$ is the average typicality of exemplars under context $e$, and the average transition probability (renormalized typicality) is $\mu^*=\frac{1}{8}$ for all the contexts. We want to identify the internal structure of state transitions of a concept making use of the typicality data. Therefore we define a transition probability (or typicality) threshold $\alpha \in [0,1]$. We say that $p\in \Sigma$  is {\it atypical} for context $e \in {\cal M}$ if and only if $\mu(p,e,\hat p) < \alpha$. The transition threshold makes it possible to express that for a given concept and a given context, there are only a limited number of possible transitions from the ground state to other states. We express this mathematically by setting the transition probability equal to zero when it is below this threshold, thereby prohibiting transitions to atypical states. Eliminating the atypical states modifies the space of possible transitions. It imposes a new renormalized transition probability distribution $\mu_\alpha$ that is limited to the transition probabilities above the threshold $\alpha$. Let 
\begin{eqnarray}
\mu^*(p,e,\hat{p},\alpha)&=&\mu(p,e,\hat{p})\quad {\rm if}\quad \mu(p,e,\hat{p})\geq \alpha \nonumber \\
&=& 0 \quad {\rm otherwise}
\end{eqnarray}
We have that $\mu_\alpha(p,e,\hat{p})=\frac{\mu(p,e,\hat{p})}{\mu^*(\alpha)}$, where $\mu^*(\alpha)=\sum_{i=1}^8\mu^*(p_i,e,\hat{p},\alpha)$.
Thus after the threshold has been imposed, a concept becomes a more constrained structure. At first glance this may appear to be an artificial bias in our analysis. However, we do not introduce the threshold to arbitrarily eliminate some exemplars, but to study the evolution of this {\it biased structure} as the threshold changes. We suggest that this may be the mechanism underlying the shift between associative and analytic thought in contextual focus. 

This leads to the next step, which is to incorporate the two modes of thought into the model by introducing a threshold-dependent notion of context relevance. The relevance of a context $e$ is associated with both the typicality of the exemplars, and the transition probabilities. When we impose a threshold $\alpha>0$, a context $e$ that has atypical exemplars may become more relevant if $\mu(p, e,\hat p)$ is higher than $\alpha$ for some specific exemplar $p$. Indeed, for each exemplar $p$ and context $e$ such that $\mu(p,e,\hat{p})>\alpha$ we have that $\mu_{\alpha}(p,e,\hat{p})>\mu(p,e,\hat{p})$. Thus, the renormalization process induced by $\alpha$ amplifies the transition probabilities of the remaining exemplars. 
In SCOP language this means the probability of considering exemplar $p$, which is atypical for the relevant contexts (and thus considered a {\it strange exemplar} for the concept) could be increased by the renormalization process if for some context $e$, $p$ is typical compared to other exemplars, i.e. $\mu(p,e,\hat{p})$ is high enough. Therefore, increasing the threshold brings some strange exemplars into play, as one might intuitively expect for associative thought. We propose that the threshold variations can be interpreted as shifts in the degree of contextual focus~\cite{Gabora2002,Gabora2003,Gabora2010}. 
\section{Analysis of the Relationship between Contexts and States}
\begin{figure}
\begin{center}
\includegraphics[width=7cm]{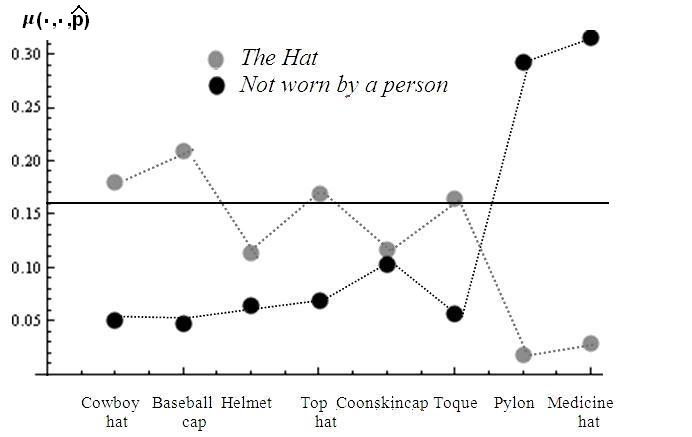}
\caption{Transition probability function of contexts {\it The hat} and {\it Not worn by a person} when $\alpha=0$, the horizontal line at $\mu(\cdot,\cdot,\hat p)=0.16$ shows the transition threshold used to identify atypical exemplars in table~\ref{data}.}
\label{prob-func}
\end{center}
\vspace{-1cm}
\end{figure}
\subsection{Context Relevance Measure}
For each context $e\in {\cal M}$ the total Likert score summed across all participants for this context $T(e)$ could be thought of as a good measure for the relevance of this context because $\frac{T(e)}{8}$ is the average typicality of the states under context $e$, and we could think {\it the higher its average typicality, the more relevant the context is}. 
A closer look reveals that this is not the case. Indeed, amongst the exemplars there might be few exemplars $p_1,...,p_n$ which are very typical for the context $e$, such that $\sum_{i=1}^n L(p_i,e)$ is almost equal to $T(e)$, but with $n$ much smaller than the total number of exemplars $N$. Thus almost all the contributions to $T(e)$ come from this small subset of exemplars. For example, consider the context $e_5$: {\it Not worn by a person} and the unit context $e_1$: {\it The hat}. Then we have $13.51=T(e_5) < T(e_1)=30.3$. But most of the contributions to $T(e_5)$ come from the exemplars $p_7$: `Toque' and $p_8$: `Medicine Hat' (note: Medicine Hat is the name of a city in Canada), because $L(p_7,e_5)+L(p_8,e_5)=8.2$. On the other hand, $T(e_1)$ is the highest of all total Likert values of all contexts because many exemplars have high Likert scores. Thus, the values of the renormalized distribution of transitions probabilities $\mu(\cdot,e_1,\hat{p})$ are spread more homogeneously amongst the exemplars. This creates a flatter distribution with smaller probability values than the more typical exemplars of the $e_5$ distribution (see figure~\ref{prob-func}), which means that it is not a very relevant context from the transition probability perspective. This observation makes it possible to explain how we can use the transition threshold to gain a clearer picture of what is going on here.

First we introduce the notion of {\it robustness} of an exemplar given a certain threshold and context. Hence, suppose we have introduced a threshold $\alpha \in [0,1]$, and consider a state $p \in \Sigma$ and a context $e \in {\cal M}$, then the robustness $R(p, e, \alpha)$ of $p$ with respect to $e$ and $\alpha$ is given by
\begin{equation}
R(p, e, \alpha)=L(p, e) \cdot \mu_{\alpha}(p, e, \hat p). 
\end{equation}
This means that the formula for state robustness compensates for both the Likert score and the renormalized transition probability induced by the threshold $\alpha$. The robustness of a state corresponds to its {\it expected typicality} in a traditional probabilistic setting. This provides a means of 
measuring
 context relevance, using our state transition probability model, and where the mode of thought lies on the spectrum of associative to analytic, determined by the threshold $\alpha$. For a given context $e$ and a given threshold $\alpha$ we define the relevance $V(e, \alpha)$ of $e$ by the expected typicality of its states: 
\begin{equation}
V(e, \alpha)=\sum_{p\in\Sigma}R(p, e, \alpha).
\end{equation}
\begin{figure}[t]
\begin{center}
\includegraphics[width=9cm]{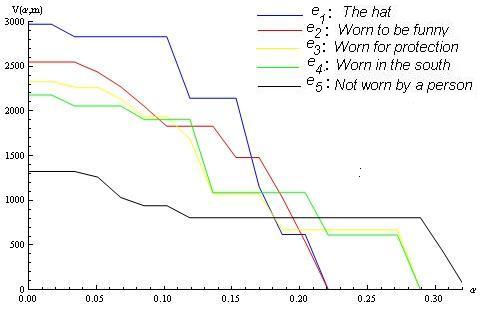}
\caption{Relevance of the contexts considered in the experiment, with respect to the threshold $\alpha$.}
\label{relevance-alpha}
\end{center}
\vspace{-0.5cm}
\end{figure}

Our results reveal an important dependency relationship between the threshold and the context relevance.  Figure~\ref{relevance-alpha} shows the function $V(e,\alpha)$ for different values of $\alpha$ for each context. Consider the contexts $e_1$: {\it The hat} and $e_5$: {\it Not worn by a person}. The curves that describe the values of the function $V(e, \alpha)$ are varying in such a way that when $0\leq\alpha\leq 0.21$, $V(e_1,0)$ is  greater than $V(e_5,0)$, and when $\alpha>0.21$, $V(e_5,\alpha)$ becomes greater than $V(e_1,\alpha)$. This means that for small thresholds, context $e_1$ is the more relevant of both, while for large thresholds, context $e_5$ is the more relevant. This is a rather unexpected situation, because context $e_5$: {\it Not worn by a person}, is at first sight not a relevant context for the concept HAT. Recall, however, that the threshold was introduced with the aim of modelling the associative mode of thought. We can also show that the other contexts exhibit a similar pattern as these two extreme ones, $e_1$, and $e_5$. For each pair of contexts, there are two possibilities. The first is that  the curve for the less relevant context at threshold $\alpha=0$ crosses the curve for the more relevant context at $\alpha=0$ for a certain threshold $\alpha>0$. (In figure~\ref{relevance-alpha} we note that this situation occurs for the pairs formed by $e_5$ with $e_4,e_3,e_2$ and $e_1$, $e_4$ with $e_2$ and $e_1$, and $e_3$ with $e_2$ and $e_1$). The second possibility is that both contexts decrease together to zero. (In figure~\ref{relevance-alpha} we note that this situation occurs for the pairs formed by $e_4$ with $e_3$, and $e_2$ with $e_1$). This suggests that we have detected the ingredients of a possible systematic pattern that can lead to the identification of rare but meaningful associations. We propose that high thresholds facilitate {\it unexpected associations} that arise, for example, in humor and creative word play, where a concept takes on a new and uncommon meaning. 
Associative thought in our framework corresponds to an exemplar $p$  and a context $e$  such that: 1)  $p$ is highly typical with respect to $e$; 2) $p$ is atypical for the set of most relevant contexts at $\alpha=0$; 3) $e$  has a low relevance compared to the set of the most relevant contexts at $\alpha=0$; 4) $e$ becomes more relevant with respect to the more relevant contexts at $\alpha=0$ for some threshold $\alpha_e>0$.
 
By increasing the typicality threshold a little above zero, we eliminate the atypical exemplars for each context. This can be interpreted as entering in a more analytic mode of thought. However, if we increase the threshold enough, we also eliminate the more typical exemplars of the most relevant contexts at $\alpha=0$ because of the {\it intrinsically} flat nature of the highly rated contexts discussed above. If the threshold is large enough, we eliminate the most common understanding of the concept, and seemingly irrelevant contexts at $\alpha=0$ become more relevant. Then some exemplars that were not so robust compared to the most robust exemplars at $\alpha=0$ become the most robust ones. These ``robust in a low-rated context'' are interpreted as the unexpected meanings that a concept can assume. This is because it is necessary to discard the most relevant meanings by defocusing attention on the most relevant contexts. This is modelled as increasing $\alpha$ to retrieve them as typical exemplars.
\begin{table}
\begin{center}
\label{Ex-typ}
\begin{tabular}{|c|c|c|c|}\hline
{ $T(e)$}    & $\#$ typical   &  Context relevance     & Type of \\ 
	                              & {\normalsize exemplars}                                 &at $\alpha=0$ &exemplar \\ \hline
Large & Large  & High & Very Representative \\
Medium & Large  & Medium & Poorly representative\\ 
{\it Medium} & {\it Small}  & {\it Low} & {\it Unexpected}\\
Small & Small & Low & Non-representative\\ \hline
\end{tabular}
\vspace{0.5cm}
\caption{Types of contexts and the type of exemplars they have.}
\end{center}
\end{table}

\section{Discussion and Future Directions}

This paper builds on previous work that uses, SCOP, a quantum-inspired theory of concepts, and psychological data, to model conceptual structure, and specifically semantic relations between the different contexts that can influence a concept. Here we focus on how these contexts come into play in analytic versus associative thought. It is suggested that the notion of a transition threshold that shifts depending on the mode of thought, as well as newly defined notions of exemplar robustness, and contextual relevance, are building blocks of a formal theory of creative thinking based on state transition probabilities in concepts. The model is consistent with the occasional finding of unexpected meanings or interpretations of concepts. The paper also strengthens previous evidence that in order to account for the multiple meanings and flexible properties that concepts can assume, it is necessary to incorporate context into the concept definition. 

The model developed here is small and idealized. In future work we plan to extend and generalize this work. An interesting parameter that we have not yet explored is the sum of the robustness of a single exemplar with respect to the set of contexts (the expected typicality of an exemplar w.r.t a set of contexts). We believe that this can be interpreted as a measure of the {\it exemplar representativeness} given in Table 2. Much as the relevance of any given context is subject to change, unexpected exemplars could become more or less representative if the transition threshold changes. Further analysis could provide a richer description of this. Another interesting development would be study the structure of the transition probabilities when  applying succesive renormailzations induced by {\it sequences of thresholds} imposed to the concept structure. We could establish, straight from the data, a threshold-dependent hierarchy of pairs $(p,e)$, that gives an account of the context-dependent {\it semantic distance} between exemplars. This could be used to model the characteristic, revealing, and sometimes surprising ways in which people make associations.

\section{Acknowledgments}

We are grateful for funding to Liane Gabora from the Social Sciences and Humanities Research Council of Canada and the Concerted Research Program of the Flemish Government of Belgium.

\bibliography{QI2011.bib}{}
\bibliographystyle{plain}
\end{document}